\documentclass[12pt]{article}
\usepackage{amsmath,amssymb,epsfig}
\usepackage{graphicx,floatflt,subfigure}
\usepackage{cite}
\usepackage{multicol}

\makeatletter

\usepackage{graphicx}
\usepackage{amssymb}
\usepackage{amsmath}
\usepackage{times}
\usepackage{latexsym}
\usepackage{multirow}
\usepackage{url}
\usepackage{color}
\usepackage{cite}
\DeclareMathOperator{\tev}{TeV}
\DeclareMathOperator{\gev}{GeV}
\DeclareMathOperator{\cm}{cm}

\def\beq{\begin{equation}}
\def\eeq{\end{equation}}
\def\bey{\begin{eqnarray}}
\def\eey{\end{eqnarray}}

\providecommand{\href}[2]{#2}

\begin{document}
\baselineskip=18pt  
\numberwithin{equation}{section}  
\allowdisplaybreaks  

\thispagestyle{empty}

\vspace*{0.8cm}
\begin{center}
 {\LARGE Theory and Phenomenology of $\mu$ in $M$ theory\\}
 \vspace*{1.5cm}
 Bobby Samir Acharya$^{1,2}$, Gordon Kane$^1$, Eric Kuflik$^1$, Ran Lu $^1$\\
 \vspace*{1.0cm}
\textsuperscript{1}{\it Michigan Center for Theoretical Physics, University of Michigan, Ann Arbor, MI 48109} \\[1ex]
\textsuperscript{2}{\it Abdus Salam International Centre for Theoretical Physics, Trieste, Italy }
 \vspace*{0.8cm}

\end{center}
\vspace*{.5cm}

\noindent

We consider a solution to the $\mu$-problem within $M$ theory on a $G_2$-manifold.
Our study is based upon the discrete symmetry proposed by Witten that forbids the $\mu$-term and 
solves the doublet-triplet splitting problem. We point out that the symmetry must be broken by moduli stabilization, describing in detail how this can occur. The  $\mu$-term is generated via Kahler interactions after strong dynamics in the hidden sector generate a potential which stabilizes all moduli and breaks supersymmetry with $m_{3/2} \sim 20 - 30 \tev$. We show that $\mu$ is suppressed relative to the gravitino mass, by higher dimensional operators, $\mu \sim 0.1 m_{3/2} \sim 2-3 \tev$. This necessarily gives a Higgsino component to the (mostly Wino) LSP, and a small but non-negligible LSP-nucleon scattering cross-section. The maximum, spin-independent cross-sections are not within reach of the current XENON100 experiment, but are within reach of upcoming runs and upgrades.

\newpage
\setcounter{page}{1}
\tableofcontents
\section{Introduction}

In the Minimal Supersymmetric Standard Model (MSSM) \cite{Martin:1997ns}, the only low energy supersymmetric parameter with mass dimension one is the $\mu$ parameter. Through the $\mu$-term, $W \supset \mu H_u H_d $, it gives mass to Higgsinos and also generates scalar potential couplings for Higgs fields. The size of $\mu$ plays an important role in phenomenology. In particular, it affects properties of potential dark matter particles. LEP searches for the charged Higgsino require $\mu \gtrsim 100$ GeV, while arguments against fine tuning of the mass of the Z-boson suggest that $\mu$ should not be too large. On the other hand, one might expect, with ignorance of the high scale theory, that  $\mu \sim m_{GUT}$, the natural UV cutoff. Solving the  $\mu$-problem \cite{Kim:1983dt}  presumably requires an understanding of the fundamental theory that generates the scale of the $\mu$ parameter. Thus the $\mu$-problem is exceptionally important--a high scale theory cannot be qualitatively complete without addressing it, and its solution will have significant implications for dark matter, Higgs physics, and fine-tuning issues.

The most promising framework for a complete fundamental theory that incorporates low energy supersymmetry is string theory. Within string theory, many explanations for the small value of $\mu$ have been proposed. In most scenarios the $\mu$-term is forbidden at the high scale. Then, it is somehow dynamically generated at a lower scale. In many cases, the $\mu$-term is forbidden by a continuous or discrete symmetry, which is spontaneously broken at a smaller, dynamically generated scale ($\ll m_{GUT}$), and perhaps related to supersymmetry breaking \cite{Antoniadis:1994hg,Nath:2002nb}. Some examples of the above include NMSSM scenarios \cite{Suematsu:1994qm,Cvetic:1995rj,Cvetic:1997ky,Lebedev:2009ag,RamosSanchez:2010sc,Ratz:2010zz} and approximate $R$-symmetric models \cite{Casas:1992mk,Kappl:2008ie}. Others scenarios have the $\mu$-term forbidden by stringy selection rules, and are broken by non-perturbative instanton effects that produce exponentially suppressed mass scales \cite{Ibanez:2007tu,Ibanez:2008my,Green:2009mx,Cvetic:2009yh}. 

It has long been suspected that the MSSM unifies the strong and electroweak forces \cite{Langacker:1991an} into a single $SU(5)$ grand unified group. Each family of quarks and leptons are organized into a $\bf{10} \oplus \bf{\bar{5}}$ representation of $SU(5)$. The remaining MSSM fields, the Higgs doublets, do not form a complete $SU(5)$ representation. Minimally, the Higgs doublets can be assigned to a  $\bf{5}\oplus \bf{\bar{5}}$ representation, but require the introduction of a pair of Higgs color triplets. The Higgs triplets can mediate baryon and lepton violating processes, and thus should be very heavy, $m_{T} \gtrsim 10^{14}$ GeV, to avoid rapid proton decay \cite{Murayama:2001ur}. Additionally, they should be heavy to ensure gauge coupling unification in the minimal model. If Higgs triplet masses are very heavy, then an $SU(5)$ symmetric theory would require that the Higgs doublets masses be the same as the triplet mass, $\mu = m_T \sim M_{GUT} $, but it was just argued that is this a factor $10^{13}$ too large. A string theoretic solution to the $\mu$-problem is inevitably related to the solution of the doublet-triplet problem of grand unified theories.

Therefore, it is paramount that the symmetry that protects the $\mu$-term not forbid the triplet masses if both problems are to be solved. This restriction leads to an elegant, perhaps unique solution to the $\mu$-problem in $M$ theory; the symmetry which protects $\mu$ from being generated at the unification scale iwas originally proposed by Witten \cite{Witten:2001bf}. Although Witten did not discuss how this symmetry would be broken, we argue that the symmetry would--indeed must-- be broken by moduli stabilization. Then by including  the mechanism for stabilizing the moduli proposed in \cite{Acharya:2006ia}, we will show that $\mu \sim 0.1\,m_{3/2}$. Finally, the implications for dark matter discovery are discussed, where we conclude that the XENON100 experiment should not observe a dark matter signal, but may do so in its next upgrade (Figure \ref{fig:xsecuniversal}).

\section{$M$ theory}

\subsection{Matter and Gauge Theory}

In $M$ theory compactified on a $G_2$ manifold, $ADE$ gauge symmetries ($SU(n)$, $SO(2n)$ and $E_6$, $E_7$, $E_8$) are localized along three dimensional submanifolds of orbifold singularities \cite{Acharya:2000gb,Acharya:1998pm}. Chiral matter, charged under the $ADE$ gauge theory, is localized at conical singularities in the seven dimensional $G_2$ manifold, at points where the $ADE$ singularity is enhanced \cite{Witten:2001uq,Acharya:2001gy,Acharya:2004qe}. Matter will additionally be charged under the $U(1)$ symmetry, corresponding to the vanishing $2$-cycle that enhances the singularity. Hence, all chiral matter will charged under at least one $U(1)$ symmetry. Bi-fundamental matter, charged under two non-Abelian gauge groups, is also possible, but will not be considered here.

As argued by \cite{Pantev:2009de}, the additional $U(1)$ symmetries are never anomalous. Therefore, there is no Green-Schwarz mechanism \cite{Green:1984sg} needed for anomaly cancellation, and GUT-scale FI $D$-terms are not present in the theory. This will be important later, since it removes a possibility for generating large scalar vacuum expectation values (vevs) for charged matter fields.

Two gauge theories will generically only have precisely the same size gauge coupling if they arise from the same orbifold singularities. Therefore, if gauge coupling unification is to be motivated theoretically, and not an approximation or accident, the gauge group of the $ADE$ singularity should be a simple group containing the Standard Model gauge group, which we will take (for simplicity) to be $SU(5)$. Any larger group containing $SU(5)$ will give results similar to those we find below. To obtain the Standard Model gauge group, $SU(5)$ needs to be broken. Perhaps the 4D gauge symmetry can be broken spontaneously, but only representations smaller than the adjoint are realizable in $M$ theory--the $\bf{10}$ and $\bf{5}$ representations (and their conjugates) in $SU(5)$. This leaves only ``flipped $SU(5)$'' \cite{Barr:1981qv,Derendinger:1983aj,Antoniadis:1987dx} as a possible mechanism to break the GUT group and solve doublet-triplet splitting. Given the difficulty in constructing a realistic flipped $SU(5)$ model \cite{Kuflik:2010dg},  it will not be considered here. The remaining possibility is to break the higher dimensional gauge theory by Wilson lines and will be discussed below.

\subsection{Moduli Stabilization}

In the mid-80's it was realized that, classically, string vacua contain a plethora of
moduli fields. The standard lore was that, after supersymmetry breaking, the moduli fields
would obtain masses and appropriate vacuum expectation values. Part of this lore was also the idea
that strong dynamics in a hidden sector would be responsible for breaking supersymmetry at, or around,
the TeV scale. Though some progress was made, 
it was not until recently that it has been clearly demonstrated that
these ideas can be completely realized in string/$M$ theory: in $M$ theory compactified on a $G_2$-manifold (without fluxes)
strong gauge dynamics can generate a potential which stabilizes all moduli and breaks supersymmetry at 
a hierarchically small scale \cite{Acharya:2006ia,Acharya:2007rc}. These vacua will be the starting point for
our considerations.

In these vacua, the gravitino mass (and therefore also the moduli masses \cite{Acharya:2010af}) $m_{3/2} \sim {\Lambda^3 \over m_{pl}^2}$, where
$\Lambda$ is the strong coupling scale of the hidden sector gauge interaction. This is parametrically of order
$\Lambda \sim e^{ -2\pi /( \alpha_{h}b)} m_{pl}$, where $\alpha_h$ is the coupling constant of the hidden sector and $b$ is
a beta-function coefficient. The vacuum expectation values of the moduli fields are also determined in terms of $\alpha_h$:
Roughly speaking, one has:
\beq
\langle s^A \rangle \sim 1/\alpha_h
\eeq
where the modulus here is dimensionless and not yet canonically normalized.
The physical meaning of the vevs of $s^A$ is that it characterizes the volumes in eleven dimensional units of 3-cycles in the
extra dimensions, e.g.,
the 3-cycle that supports the hidden sector gauge group. Thus, self-consistently when the hidden sector is weakly coupled
in the UV, the moduli are stabilized at large enough volumes in order to trust the supergravity potential which only makes sense in
this regime.  
In general, the rough formula exhibits the scaling with $\alpha_h$ and, numerically the moduli vevs in the vacua considered thus far range from about $1 \leq s^A \leq 5/\alpha_h$.

In order to incorporate the moduli vevs into the effective field theory in an $M$ theory vacuum, we have
to consider the normalized dimensionful vevs which appear in the Einstein frame supergravity Lagrangian.
For obtaining the normalization it suffices to consider the moduli kinetic terms alone:
\beq
\mathcal{L} \supset m_{pl}^2 {1\over 2} g_{AB} \partial_{\mu} s^{A} \partial^{\mu} s^{B} \label{modulikin}
\eeq
where $s^A$ are the dimensionless moduli described above and $g_{AB}$ is the (Kahler) metric on
the moduli space. From the fact that the extra dimensions have holonomy $G_2$, it follows that
each component of $g_{AB}$ is homogeneous of degree {\it minus} two in the moduli fields
\beq
g_{AB} = \partial_A \partial_B K = \partial_A \partial_B\left(-3 \ln V_7+ \dots \right) 
\eeq
because the volume of $X$, $V_7$ is homogeneous of degree $7/3$.

For isotropic $G_2$-manifolds, i.e. those which receive similar order contributions to their
volume from each of the $N$ moduli, studying examples shows that, not only is the
metric of order ${1 \over s^2}$, but also of order $1/N$:
\beq
 g \sim \frac{1}{N} \frac{1}{(s^A)^{2}}
\eeq 

Therefore in a given vacuum the order of magnitude of the entries of $g_{AB}$ are
\beq
g \sim \frac{\alpha_h^2}{N} 
\eeq

Therefore, a dimensionless modulus vev of order $1/\alpha_h$ translates into a properly normalized dimensionful vev 
\beq 
\langle \hat{s^A} \rangle \sim \frac{1}{\sqrt{N}} \sim 0.1 m_{pl}
\eeq
for $N\sim 100$, which is a typical expectation for the number of moduli \cite{joyce}\footnote{Presumably, $N$ is of the same order as
the number of renormalizable coupling constants of the effective low energy theory.}. 

This can lead to a suppression of the effective couplings which generate the $\mu$-term, once the symmetry forbidding $\mu$ is broken.   More precise calculations of the moduli vevs can be found in \cite{Acharya:2007rc,Acharya:2008zi}. Clearly, however, a $G_2$-manifold with less than ten or so moduli will not have 
suppressed, normalized moduli vevs; such cases are presumably unlikely candidates for $G_2$-manifolds with realistic particle spectra and will
not be considered further.

We briefly also discuss the spectrum of Beyond Standard Model (BSM) particles which arise from the $M$ theory vacuum. Classically, it is well known that
string/$M$ theory has no vacuum with a positive cosmological constant (de Sitter minimum). From the effective field theory
point of view, this is the statement that moduli fields tend to have potentials which, in the classical limit have no
de Sitter minimum. If we now consider quantum corrections to the moduli potential, which {\it only} involve the moduli fields -- if they are computed in a perturbative regime -- they tend to be small and hence are unlikely to generate de Sitter vacua.
Positive, larger sources of vacuum energy must therefore arise from other, non-moduli fields. This is indeed the case in the
$M$ theory vacua described in \cite{Acharya:2007rc}. Here the dominant contribution to the vacuum energy arises from a {\it matter}
field in the hidden sector (where it can be shown that, without the matter field, no de Sitter vacuum exists). This is important
for the following reasons.

Adopting supersymmetric terminology, this suggests that the fields with the dominant $F$-terms are not moduli.
Hence, the moduli $F$-terms are suppressed relative to the dominant contribution (in fact, in $M$ theory the suppression is of 
order $\alpha_h$). This affects the spectrum of BSM particles. In string/$M$ theory, gaugino masses are generated through $F$-terms
of moduli vevs (because the gauge coupling function is a superfield containing volume moduli). Hence, at leading order these will
be suppressed relative to, say, scalar masses which receive order $m_{3/2}$ contributions from all $F$-terms in the absence of accidental
symmetries. Therefore, in the $G_2$-MSSM (and presumably other classes of string vacua) the scalar superpartners and moduli fields will
have masses of order $m_{3/2}$ whereas the gaugino's will have masses which are suppressed; in fact in the $G_2$-MSSM the gaugino masses
at the GUT scale are
at least two orders of magnitude below $m_{3/2}$. This is what
makes the anomaly mediated contributions to gaugino masses relevant to the $G_2$-MSSM and also why the models often contain a Wino
LSP \cite{Acharya:2008zi}. Important for our considerations below will be the fact that the suppression of the gaugino masses
is greater than the suppression of moduli vevs discussed above by one order of magnitude (at the GUT scale), at least for $G_2$-manifolds with less
than O($10^4$) moduli.

\subsection{Geometric Symmetries and Moduli Transformations}

Compact, Ricci-flat manifolds with finite fundamental groups, such as manifolds with holonomy $G_2$
or $SU(3)$ can not have continuous symmetries. They can, however, have {\it discrete} symmetries.
Witten was considering just such a discrete symmetry ($G$) of a $G_2$-manifold when he proposed the symmetry
which prevents $\mu$. Assuming the simplest possibility of an Abelian discrete symmetry, let us consider $G = {\bf Z_N}$, which acts on $X$:
\beq
{\bf Z_N}: X \longrightarrow X
\eeq
As a result of this, it will also act naturally on the fields on $X$. In particular ${\bf Z_N}$ will act on the set of
harmonic forms on $X$. Our interest here is $H^3 (X, {\bf R})$ the set of harmonic 3-forms on $X$, since this locally
represents the moduli space of $G_2$-manifolds. A $G_2$-manifold, with moduli at a point $\langle s^S \rangle = s_0^A$ is determined by a harmonic (locally) $G_2$ invariant 3-form 
$\varphi$ as
\beq
\varphi = \sum s_0^A \beta_A
\eeq
where $\beta_A$ are a basis for $H^3 (X, {\bf R})$. If the point $s^A_0$ is such that ${\bf Z_N}$ is a symmetry, then $\varphi$ will be invariant
under ${\bf Z_N}$, because invariance of $\varphi$ is equivalent to invariance of the metric. The three-forms $\beta_A$  transform in a representation of ${\bf Z_N}$, which is a real representation because the 3-forms are real on
a $G_2$-manifold. Hence,
\beq
{\bf Z_N}:\;\; \beta_A \rightarrow M_A^B\;\beta_B
\eeq
where $M$ is defined by this equation.

The fact that the particular $G_2$-manifold, characterized by the particular point in moduli space $s_0^A$, is ${\bf Z_N}$-invariant is simply the statement that:
\beq
s_0^B M_A^B = s_0^A
\eeq
i.e., the $s_0^A$ are an eigenvector of $M$ with unit eigenvalue. Clearly, this will not be true for a generic vector $s^A$; hence, for a generic
point in the moduli space, the entire ${\bf Z_N}$ symmetry will be broken. Since the representation of ${\bf Z_N}$ defined by the matrix $M$ is
real, it must be the sum of a complex representation plus its conjugate.
Thus, the basis $\beta_B$ can be chosen such that the complex representation is spanned by {\it complex} linear combinations of moduli fields.
For instance, there might be a linear combination 
\beq
S = \hat{s}^1 + i \hat{s}^2 \label{complexmoduli}
\eeq
which we choose to write in-terms of the dimensionful fields ($\hat{s}$), that transforms as
\beq
S \rightarrow e^{2\pi i/N} S.
\eeq

Since we usually consider complex representations of discrete symmetries acting on the matter fields in effective field theories, it will be
precisely the linear combinations of moduli (those in the form (\ref{complexmoduli})) which span ${\bf r_C}$ which will appear in the "symmetry breaking sector" of the effective Lagrangian. In other words, the moduli will appear in complex linear comibinations such as (\ref{complexmoduli}) in the Kahler potential operators containing other fields that transform under the ${\bf Z_N}$. Note that in (\ref{complexmoduli}) we are abusing notation in the sense that the "$i$" which
appears is in general an $N$-by-$N$ matrix whose square is minus the identity.

\section{Witten's Solution}

In heterotic and type-II string theories doublet-triplet splitting is often solved via orbifold compactifications \cite{Hosotani:1983vn,Witten:1985xc}. In these theories, higher (space-time) dimensional gauge symmetries are broken by the Wilson lines in an orbifold compactification, while the Kaluza-Klein zero mode Higgs triplets are absent due to non-trivial transformations under the orbifold symmetry. On the contrary, matter fields in $M$ theory are co-dimension 7, that is, the fields live only in four dimensions, and are not zero modes of a KK tower of fields, so this solution to the $\mu$-problem will not work. Other possibilities, such as NMSSM realizations or string instanton effects, will also not work since the symmetry that forbids $\mu$ (a $U(1)$ or stringy selection rules) would also forbid the triplet mass, thus spoiling doublet-triplet splitting.

One may also consider the possibility that a discrete $R$-symmetry can forbid the $\mu$-term while solving doublet-triplet splitting. Requiring the symmetry to be anomaly free, and that it commutes with the gauge theory can lead to a unique symmetry \cite{Lee:2010gv}. However, this symmetry will also forbid the triplet mass and spoil doublet triplet splitting unless the triplets are absent from the  four dimensional theory. For most string theories, this can be accomplished by a Wilson line in the higher dimensional theory, but in $M$ theory, this is not possible since matter only exists in four dimensions. 

Therefore, an alternative approach is needed to solve doublet-triplet splitting in $M$ theory. The only known possibility, originally discussed by Witten, is to construct a discrete $ {\bf Z_N}$ symmetry of the geometry, that will act on both matter fields and moduli-fields. When combined with a discrete Wilson line thats breaks $SU(5)$, this symmetry need not commute with the $SU(5)$, thus allowing components of a single $SU(5)$ representation to have different ${\bf Z_N}$ charges. Since the above arguments demonstrate that there must be a symmetry that acts differently on doublets and triplets, so far this is the only approach known to work, and maybe be the only solution. 

The minimal $SU(5)$ matter content contains three generations of matter descending from three copies of $\bf{10}_M \oplus \bf{\bar{5}}_M$. There is also a $\bf{5}_H \oplus \bf{\bar{5}}_{H}$ pair containing the MSSM Higgs doublets, $H_u \oplus H_d$, and a vector-like pair of Higgs triplets, $T_u \oplus T_d$. Here a doublet and a triplet from one of the Higgs representations can transform differently under the ${\bf Z_N}$ symmetry group. Without loss of generality or phenomenology, this field is taken to be the $\bf{\bar{5}}_{H}$ field, with the following charges for the fields
\begin{equation}\begin{array}{rc|c}
\multicolumn{2}{l}{\mbox{Field}} & {\bf Z_N}  \\ \hline

\multicolumn{2}{c|}{ \mathbf{10}_M } &  \eta^\sigma \\
\multicolumn{2}{c|}{ \mathbf{\overline{5}}_M  } &   \eta^\tau   \\
\multirow{2}{*}{$\mathbf{5}_H  $} & T_u & \eta^\alpha \\
 & H_u & \eta^\alpha \\
\multirow{2}{*}{$\mathbf{\overline{5}}_{\overline{H}} $} & T_d & \eta^\gamma \\
 & H_d & \eta^\delta
\end{array}\end{equation}
where $\eta\equiv e^{2\pi i / N}$.

These charges are constrained by the requirement that the $\bf Z_N$-symmetry does not forbid necessary terms in the superpotential, such as Yukawa couplings, Majorana neutrino masses, and the Higgs triplet masses
\begin{equation}\begin{array}{rc|c}
\multicolumn{2}{c}{\mbox{Coupling}\;\;\;\;\;}  &   \mbox{Constraint }\\ \hline
\mbox{Up Yukawa Coupling} & \mathbf{10}_M \mathbf{10}_M H_u & 2 \sigma + \alpha  = 0 \mod N\\
\mbox{Down Yukawa Coupling} & \mathbf{10}_M \mathbf{\overline{5}}_M H_d & \sigma + \tau + \delta = 0 \mod N\\
\mbox{Majorana Neutrino Masses} & H_d H_d \mathbf{\overline{5}}_M \mathbf{\overline{5}}_M & 2 \alpha + 2 \tau = 0 \mod N\\
\mbox{Triplet Masses} & T_u T_d & \alpha + \gamma= 0 \mod N.\\
\end{array}\end{equation}
The solution to these equations are
\begin{equation}\begin{array}{ccl}
\alpha &=& -2 \sigma \\
\gamma &=& 2 \sigma  \\
\delta &=& -3 \sigma + N/2  \\
\tau   &=& 2 \sigma  + N/2  \\
\sigma &=& \sigma.
\end{array}\end{equation}
Inherently, the $\bf Z_N$ should forbid the $\mu$-term, and if possible, other dangerous terms, such as dimension-5 proton decay operators and dimensions 3 and 4 R-parity violation.
\begin{equation}\begin{array}{rc|c}
\multicolumn{2}{c}{\mbox{Coupling}}  &   \mbox{Constraint }\\ \hline
\mu-\mbox{term} 	& H_d H_u 							     & -5 \sigma + N/2 \ne 0 \mod N\\
\mbox{D-5 Proton Decay} & \mathbf{10}_M \mathbf{10}_M \mathbf{10}_M \mathbf{\overline{5}}_M  & 5\sigma - N/2 \ne 0 \mod N\\
\mbox{D-3 R-Parity} & \mathbf{5}_H \mathbf{\overline{5}}_M 			             & N/2 \ne  0 \mod N\\
\mbox{D-4 R-Parity} & \mathbf{10}_M \mathbf{\overline{5}}_M \mathbf{\overline{5}}_M          & 5\sigma  \ne 0 \mod N.\\
\end{array}\end{equation}
Doublet-triplet splitting occurs if $5 \sigma \ne N/2 \mod N $. If one only wants to solve  doublet-triplet splitting while forbidding the $\mu$-term, then there is a solution for $N=2$ and $\sigma=1$. Forbidding all the dangerous operators above can be accomplished with a ${\bf Z_4}$ symmetry.

An essential point is that the existing bounds coming from the LEP experiments assert that the masses of charged Higgsinos are at least 100 GeV, hence an effective 
$\mu$-term must be generated. In our context here this implies that the ${\bf Z_N}$ symmetry must be broken, an aspect
not discussed in \cite{Witten:2001bf}. 
This symmetry
breaking is the subject of the next section.

\section{Generating $\mu$}

As discussed in above, the ${\bf Z_N}$ symmetry is a geometric symmetry of the internal $G_2$ manifold, under which the moduli are charged. 
The $G_2$ moduli \cite{Acharya:2006ia} reside in chiral supermultiplets whose complex scalar components, 
\beq z^j =t^j + i s^j ,\eeq
are formed from the geometric moduli of the manifold\footnote{Note section 2. The "$i$"'s are not the same in $S$ and $z$.}, $s_i$, and axionic components of the three-from $C$-field, $t_i$. We expect   
the moduli to break the discrete symmetry just below Planck scale when their vevs are stabilized \cite{Acharya:2007rc,Acharya:2008zi} (see Section (2.2)),
\beq 
\left< \hat{s}_i \right> \sim 0.1 m_p .  \label{modulivev}
\eeq
Likewise, the moduli $F$ terms are expected to give gaugino masses in the usual way, so that 
\beq  
\left< F_{z^i} \right> \simeq m_{1/2} \, m_p  \label{moduliFvev} .
 \eeq
where $m_{1/2}$ is the tree level gaugino mass at the GUT scale.
The axion shift symmetries $t_i \rightarrow t_i + a_i $ require that only imaginary parts of the moduli appear in perturbative interactions. The superpotential, being holomorphic in the fields, will not contain polynomial terms that explicitly depend on the moduli.  The $\mu$-term can then only be generated via Kahler interactions when supersymmetry is broken via a Guidice-Massiero like mechanism \cite{Giudice:1988yz}, i.e., from Kahler potential couplings quadratic in the Higgs fields. 

To understand the size of $\mu$ (and $B\!\mu$) we we first find a combination of moduli fields (or product of moduli fields), invariant under the axion symmetries, that transform under (a complex representation of ) $\bf Z_N$ with charge $5 \sigma - N/2$
\beq
S^1 = \hat{s}^{i} + i \hat{s}^{j} \label{S1}
\eeq
and another with charge $-5 \sigma - N/2$
\beq
S^2 = \hat{s}^{m} + i \hat{s}^{n} \label{S2}  .
\eeq
These fields have the correct charge to break the symmetry and generate the $\mu$-term which has total ${\bf Z_N} $ charge $-5 \sigma - N/2$.

In a general supergravity theory \cite{Wess:1992cp,Brignole:1997dp} the fermion mass matrix is
\beq 
m^\psi_{ij} = m_{pl}^3 e^{G/2} \left( \nabla_i G_j +  G_i G_j \right) \label{fermmasses}
 \eeq
and the holormorphic components of the scalar mass matrix are
\beq 
m^{\phi~ 2}_{ij}  = m_{pl}^4 e^{G} \left( \nabla_i G_j + G^k \nabla_i \nabla_j G_k \right) \label{scalarmasses}
 \eeq
where $G = m_{pl}^{-2} K +  \ln (m_{pl}^{-6} |W|^2)$ and subscripts on $G$ denote derivatives with respect to the scalar fields $\phi_i$ or their conjugates $\phi^{*}_{\bar{i}}$. Respectively, (\ref{fermmasses}) and (\ref{scalarmasses}) can be used to find $\mu$
\beq 
 \mu =  \langle m_{3/2} K_{\mbox{\tiny{$H_u H_d$}}} - F^{\bar{k}} K_{\mbox{\tiny{$H_u H_d$}}\bar{k}}  \rangle 
 \eeq
and $B\mu$
\beq \begin{array}{ccl}
B\!\mu &=&   \langle 2 m_{3/2}^2 {K}_{\mbox{\tiny{$H_u H_d$}}} -    m_{3/2}F^{\overline{k}} {K}_{\mbox{\tiny{$H_u H_d$}}\bar{k}} + m_{3/2} F^m  K_{\mbox{\tiny{$H_u H_d$}}m } \\
&-&    \left( m_{3/2} F^m
 K^{n \overline{l}} K_{\overline{l} m \mbox{\tiny{$H_u$}} } K_{n \mbox{\tiny{$H_d$}}} +(\mbox{\tiny{$H_d$}} \leftrightarrow \mbox{\tiny{$H_u$}})\right) \\
&-& F^n F^{\bar{m}}  \left( \frac{1}{2}    K_{\mbox{\tiny{$H_u H_d$}}n\bar{m}}  -  
 K^{j \overline{l}} K_{\overline{l} n \mbox{\tiny{$H_u$}} } K_{j \bar{m} \mbox{\tiny{$H_d$}}} +(\mbox{\tiny{$H_d$}} \leftrightarrow \mbox{\tiny{$H_u$}})\right) \rangle
\end{array} \eeq
where the indices run over the moduli fields and we have used that the superpotential does not contribute to either mass. Leading contributions come from Kahler potential terms
\beq
 K \supset \alpha \frac{{(S^1)}^{\dagger}}{m_{pl}} H_u H_d +  \beta \frac{{(S^2)}}{m_{pl}} H_u H_d + h.c. \label{Kahlermu}
\eeq
where the coefficients $\alpha,\beta$ are expected to be $\mathcal{O}(1)$. Plugging the Kahler potential ( \ref{Kahlermu}) into the formulas for $\mu$ and $B\!\mu$ gives the $\mu$-term
\beq \begin{array}{ccl}
 \mu &=&  \alpha \frac{\langle{S^1} \rangle}{m_{pl}}  m_{3/2} + \alpha  \frac{ \langle F^{S^1} \rangle }{m_{pl}} \\
B\!\mu &=&  2 \alpha \frac{\langle {S^1} \rangle}{m_{pl}}  m_{3/2}^2 + \alpha  \frac{ \langle F^{{S^1}} \rangle }{m_{pl}} m_{3/2} + \beta  \frac{ \langle F^{{S^2}} \rangle }{m_{pl}} m_{3/2}. 
\label{muterm}
\end{array}\eeq
However, as a result of (\ref{modulivev}), (\ref{moduliFvev}) and the suppression of $m_{1/2}$ by order two orders of magnitude in the $G_2$-MSSM,  $\langle S^i \rangle   m_{3/2} \simeq 10 \;\;\langle F^{S^i} \rangle $, the contribution to the masses coming from $F$-terms are sub-dominant, at least if we assume
that $N\ll10^4$.
Therefore, to a good approximation
\beq
B\!\mu \simeq 2\, \mu m_{3/2} 
\eeq
a fact which will have significant phenomenological consequences\footnote{We leave the case of $N \geq 10^4$ for further study.} . 

The coefficients of the operators in (\ref{Kahlermu}) are in principle determined from $M$ theory, but is not yet known how to calculate them precisely.  It is natural
to assume that the 
 coupling coefficients are of $\mathcal{O}(1)$.  When combined with a model of moduli stabilization, such as in the $G_2$-MSSM described in \cite{Acharya:2006ia,Acharya:2007rc,Acharya:2008zi} and briefly reviewed section (2.2),  $\mu$ and $B\!\mu$ can be approximately determined. Since the real and imaginary components of the complex fields, $S^1$ (\ref{S1}) and $S^2$ (\ref{S2}), are expected to have similar, but not necessarily identical vevs, $\mu$ will generically have a phase, that will be unrelated to the phases that enter the gaugino masses. But, $B\mu$ and $\mu$ will have the same phase since both are proportional to $S^1$ and the same coupling constant.

Before moving on to the next section we discuss the possibility that other matter fields may be charged under the ${\bf Z_N}$ symmetry, spontaneously break the ${\bf Z_N}$ symmetry, and generate $\mu$. Consider an $SU(5)$-singlet matter field $X$ that generates the $\mu$-term via the superpotential coupling $X H_u H_d$. Since $X$ is a matter field, $M$ theory requires that it is charged under least one $U(1)$ symmetry. Then $H_u H_d$ is not invariant under the $U(1)$, and consequently,  the triplet mass term $T_d T_u$ is not invariant, spoiling doublet-triplet splitting. Thus, such contributions should not occur.

Alternatively, the $\mu$-term may be generated by a $U(1)$ invariant combination of two fields, for example by the operator
\beq
\frac{X_1 X_2}{\Lambda} H_u H_d.
\eeq
 Requiring $\mu \gtrsim 10^3$ GeV, and taking $\Lambda \sim M_{GUT}$ this would require $\sqrt{ \langle X_1 X_2 \rangle } \gtrsim 10^{10}$ GeV. Radiative symmetry breaking will generally give a vev $\sim m_{3/2}$-- usually large vevs are associated with FI $D$-terms. But since FI $D$-terms are absent in $M$ theory, it may be difficult for such large vevs to arise from here. The recent results of \cite{Acharya:2011kz} do suggest that the $F$-term potential can generate large
matter field vevs, however in that case the vevs are too large to be relevant for the $\mu$ problem.
Therefore, we very tentatively conclude that a matter field spurion is not responsible for breaking the ${\bf Z_N}$ symmetry and giving a physically relevant $\mu$-term.

Finally, we comment on a potential domain wall problem. The moduli are stabilized away from a  ${\bf Z_N}$ point, which implies that the ${\bf Z_N}$ symmetry was really only an approximate symmetry of the $G_2$-manifold.  The moduli stabilization serves to parameterize the amount that the $G_2$-manifold differs from a ${\bf Z_N}$ symmetric manifold. Therefore, since the ${\bf Z_N}$ symmetry is not an exact symmetry of the $G_2$ manifold,  the Lagrangian will explicitly break the ${\bf Z_N}$ symmetry, and domains walls would not have formed in the early universe.

\section{Origin of $R$-Parity in $M$ theory}

In the Standard Model, the Yukawa couplings and Higgs potential form the most general set of renormalizable couplings consistent with the gauge symmetries. In this sense, baryon (B) and lepton (L) number are accidental symmetries of the theory. However, this is not the case in supersymmetric theories, which allow for the B and L violating renormalizable couplings\footnote{ The final term in (\ref{WRbreaking}) can be rotated away in superpotential by a unitary transformation on $(h_d, L)$. This rotation will induce additional contributions to the lepton violating coupling constants $\lambda^{\prime}$ and $\lambda^{\prime\prime}$ that are proportional to the Yukawa couplings. Assuming that $\mu \gtrsim \kappa $, their sizes are approximately 
\beq 
\lambda^{\prime} \sim y_e \frac{\kappa}{\mu} \;\;\;\;\;\; \lambda^{\prime\prime} \sim y_d \frac{\kappa}{\mu} \label{muyukawa}
\eeq
}
\beq
 W_{\not{R}} = \lambda^{\prime} L L e^c + \lambda^{\prime\prime} L Q d^c + \lambda^{\prime \prime\prime} u^c d^c d^c + \kappa L h_u .
\label{WRbreaking}
\eeq
If the squark masses are not of order the GUT scale (which presumably they are not), these operators can lead to too rapid proton decay if not heavily suppressed. Hence one usually introduces $R$-parity, where the Standard Model fields have $R$-parity $+1$, while their supersymmetric partners have $R$-parity $-1$. This forbids all the couplings in   (\ref{WRbreaking}).

Additionally, $R$-parity invariance insures the stability of the LSP, and the absence of an $R$-parity can eliminate the LSP as a dark matter candidate.  Therefore, in this section we will discuss the origin of $R$-parity in $M$ theory, or at least an approximate $R$-parity that leaves the proton and LSP very long lived. Of course from a theoretical point of view an $R$-parity or equivalent symmetry should emerge from the theory and not be put in by hand.

The ${\bf Z_N}$ symmetry constructed in Section 3 contains $R$-parity, but for generic moduli charges the complete ${\bf Z_N}$ symmetry, including any $R$-parity subgroup, will be spontaneously broken. Although the ${\bf Z_N}$ symmetry will prevent the superpotential couplings in (\ref{WRbreaking}) from being  invariant, supersymmetry breaking will revitalize the operators just as in the case of the $\mu$-term, from Kahler potential operators
\beq
 K_{\not{R}} \supset \frac{\tilde{S}^{ \dagger}}{m^2_{pl}} L L e^c + \frac{\tilde{S}^{ \dagger}}{m^2_{pl}}  L Q d^c + \frac{\tilde{S}^{ \dagger}}{m^2_{pl}}  u^c d^c d^c + \frac{\tilde{S}^{ \dagger}}{m_{pl}} L h_u \label{kahlerbad1}
\eeq
where the $\tilde{S}^{ \dagger}$'s symbolically represent the moduli and need not all be the same.

Just as the $\mu$-term was generated from the Kahler potential as a result of moduli stabilization, the effective superpotential can be calculated from the supersymmetry breaking contribution from (\ref{kahlerbad1}) to 
\beq \begin{array}{ccl}
\lambda_{i j k} &\simeq& {m_{pl}^{-2}} (\langle \tilde{S} \rangle m_{3/2} +F_{\tilde{S}} )( K_{\not{R}} )_{i jk} \\
\kappa &\simeq& m_{pl}^{-1} (\langle \tilde{S} \rangle m_{3/2} +F_{\tilde{S}})( K_{\not{R}} )_{L h_u} \label{kappalambda}
 \end{array} \eeq
for $\lambda = \lambda^{\prime}, \lambda^{\prime\prime}, \lambda^{\prime\prime\prime}$ and where $ i,j,k$ run over the matter fields.  Comparing (\ref{kappalambda}) to (\ref{muterm}), one easily sees that $\kappa \sim  \mu$, since both are generated the same way.

Then using that $\mu \sim  \kappa $,  the superpotential can be rewritten as
\beq
 W_{\not{R}} \simeq  \frac{\mu }{m_{pl}} (L L e^c +  L Q d^c  +  u^c d^c d^c )+  \mu  L h_u.
\eeq
The trilinear couplings are suppressed but the lepton violating bilinear coupling is large and of order the $\mu$-term--this is simply a consequence of $\kappa$ not being suppressed. After rotating away the $L h_u$ term using the approximation (\ref{muyukawa}), the superpotential simplifies to
\beq
 W_{\not{R}} \sim y_e L L e^c +  y_d L Q d^c  +  \frac{\mu}{m_p} u^c d^c d^c
\label{WRbad1} 
\eeq
where smaller terms in $\lambda^{\prime},\lambda^{\prime\prime},\lambda^{\prime\prime\prime}$ have been dropped. Thus the  lepton number violating trilinears pick up large contributions from the bilinear term, even if they were originally suppressed.

The proton lifetime for the decay mode $p \rightarrow e^{+} \pi^0$ is estimated to be
\beq
\Gamma_{p \rightarrow e^{+} \pi^0} \simeq \frac{\lambda^{\prime\prime 2}}{4 \pi} \frac{\lambda^{\prime \prime \prime 2}}{4 \pi}  
\frac{m_{\text{proton}}^5}{m_0^4}.
\eeq
The current bounds on this partial decay width is  $\tau_{p \rightarrow e^{+} \pi^0} > 1.6 \times 10^{33}$ years \cite{Amsler:2008zzb}. For scalar masses in the $G_2$-MSSM ($\sim 10 \tev$ see \cite{Acharya:2008zi}) this gives the experimental bound
\beq
 \lambda^{\prime\prime } \lambda^{\prime \prime \prime} \lesssim 10^{-24} 
\eeq
which clearly excludes the superpotential (\ref{WRbad1}), since $ \lambda^{\prime\prime } \sim y_e \sim 10^{-5}$ and  $ \lambda^{\prime\prime \prime} \sim \mu / m_{pl} \sim 10^{-14}$. Therefore, proton stability requires an additional form of $R$-parity invariance beyond the discrete symmetry proposed.

One possible way to preserve the $R$-parity is to simply assume that the $G_2$-manifold in the vacuum is $R$-parity invariant, though
not ${\bf Z_N}$ invariant i.e. the vacuum partially breaks ${\bf Z_N}$ to an $R$-parity subgroup.
For example, take $N=6$, then
\beq\begin{array}{rc|c}
\multicolumn{2}{r}{\mbox{Coupling}}   &{\bf Z_6}\mbox{ charge} \\ \hline
\mu-\mbox{term} 	& H_d H_u 							   &  \eta^4\\
\multirow{2}{*}{\mbox{R-Parity}} & M_{10} M_{\bar{5}} M_{\bar{5}}     & \eta^5 \\
& M_{\bar{5}} H_u      &  \eta^3
\end{array}\eeq
for $\eta \equiv  e^{i 2 \pi /6}$. If all moduli transform under the ${\bf Z_3}$ subgroup of ${\bf Z_6}$, then ${\bf Z_6}$ is broken to ${\bf Z_2}$ $R$-Parity, since no $R$-parity couplings can be generated. This is technically satisfactory, but is presumably "non-generic". It could certainly emerge from $M$ theory, 
but we will not consider it further here.

Alternatively,  $R$-parity may manifest itself as matter-parity, a conserved remnant of a local, continuous $U(1)$ symmetry. As is well known, matter parity arises naturally in $SO(10)$ theories. When embedded into an $SO(10)$ unified theory, the Standard Model matter fields belong to a different representation than the Higgs fields-- a generation of matter is contained in a $\bf{16}$ of $SO(10)$, while a pair of Higgs doublets comes from a $\bf{10}$ of $SO(10)$. 

When $SO(10)$ is broken to $SU(5)\times U(1)_{\chi}$, for example by a discrete Wilson line, the Higgs fields and matter fields are charged differently under $U(1)_{\chi}$:
\beq\begin{array}{ccl}
 SO(10) & \rightarrow & SU(5)\times U(1)_{\chi} \\
\bf{16} & \rightarrow & \bf{10}_{-1} \oplus \bf{\bar{5}}_{3} \oplus \bf{1}_{-5} \\
\bf{10} & \rightarrow & \bf{5}_{2} \oplus \bf{\bar{5}}_{-2}.
\end{array}\eeq
where the subscript is the $U(1)_{\chi}$ charge.

The vacuum expectation values of the Higgses, which are contained in the $\bf{5}_{2}$ and  $\bf{\bar{5}}_{-2}$ multiplets, will break the $U(1)_{\chi}$ symmetry into a discrete ${\bf Z_2}$ subgroup. This is because the Lagrangian is no longer invariant under the full local transformation $\Phi \rightarrow e^{i \alpha(x) q_d } \Phi$, but only the subgroup of transformations given by $\alpha(x)=\pi$. In terms of the $U(1)_{\chi}$ charges $q_{\chi}$, the chiral multiplets have ${\bf Z_2}$-parity $e^{i \pi q_{\chi}}$. Thus chiral superfields with even $U(1)_{\chi}$ charge will have parity $+1$ and fields with odd $U(1)_{\chi}$ charge will have parity $-1$. The ${\bf Z_2}$ symmetry is exactly $R$-parity.

The only $SU(5)$ singlet with $U(1)_{\chi}$ charge is the $\bf{1}_{-5}$ field (and its conjugate), and thus this is the only field that can break $U(1)_{\chi}$ without breaking the SM gauge group. But since it has odd $U(1)_{\chi}$ charge, its vev will break $R$-parity. Therefore an $SO(10)$ completion of $U(1)_{\chi}$ will not contain an unbroken $R$-parity, but perhaps when combined with the ${\bf Z_N}$ symmetry, $R$-parity violating operators may be sufficiently suppressed to allow a long lived proton and LSP. Next we estimate these lifetimes. 

The singlet field $\bf{1}_{-5}$ can be considered to be the right-handed neutrino, $\nu^c$, since it has the right quantum numbers to make the operator
$ \nu^c h_u L$ invariant under $U(1)_{\chi}$. However, if $\langle \nu^c \rangle \ne 0 $, all baryon and lepton violating operators in  (\ref{WRbreaking}) will be generated via the superpotential
\beq
 W_{\not{R}} \sim \nu^c L L e^c +  \nu^c L Q d^c  +  \nu^c u^c d^c d^c + \nu^c  h_u L .
\label{WRbad2}
\eeq
The operators in (\ref{WRbad2}) should be suppressed and can be forbidden by the ${\bf Z_N}$ symmetry. The story will be the same as above and the Kahler potential operators will generate (\ref{WRbad2}) , but with additional suppression coming from $U(1)_{\chi}$ breaking
\beq
 W_{\not{R}} = (\frac{\langle \tilde{S} \rangle m_{3/2} +F_{\tilde{S}} }{m_{pl}^2})( \frac{\langle \nu^c \rangle}{m_p}) (L L e^c +  L Q d^c +  u^c d^c d^c )+ (\frac{\langle \tilde{S} \rangle m_{3/2} +F_{\tilde{S}} }{m_{pl}}) ( \frac{\langle \nu^c \rangle}{m_p}) L h_u.
\eeq
Diagonalizing away the $L h_u $ term, and using (\ref{muterm}) gives
\beq
 W_{\not{R}} \sim y_e \frac{\langle \nu^c \rangle}{m_p} L L e^c +  y_d\frac{\langle \nu^c \rangle}{m_p} L Q d^c  +  \frac{\mu}{m_p} \frac{\langle \nu^c \rangle}{m_p} u^c d^c d^c. 
\label{WRbadc} 
\eeq
where again large lepton violating trilinear terms are induced by the rotation. 

\begin{figure}
\centering
\includegraphics[width=1\textwidth]{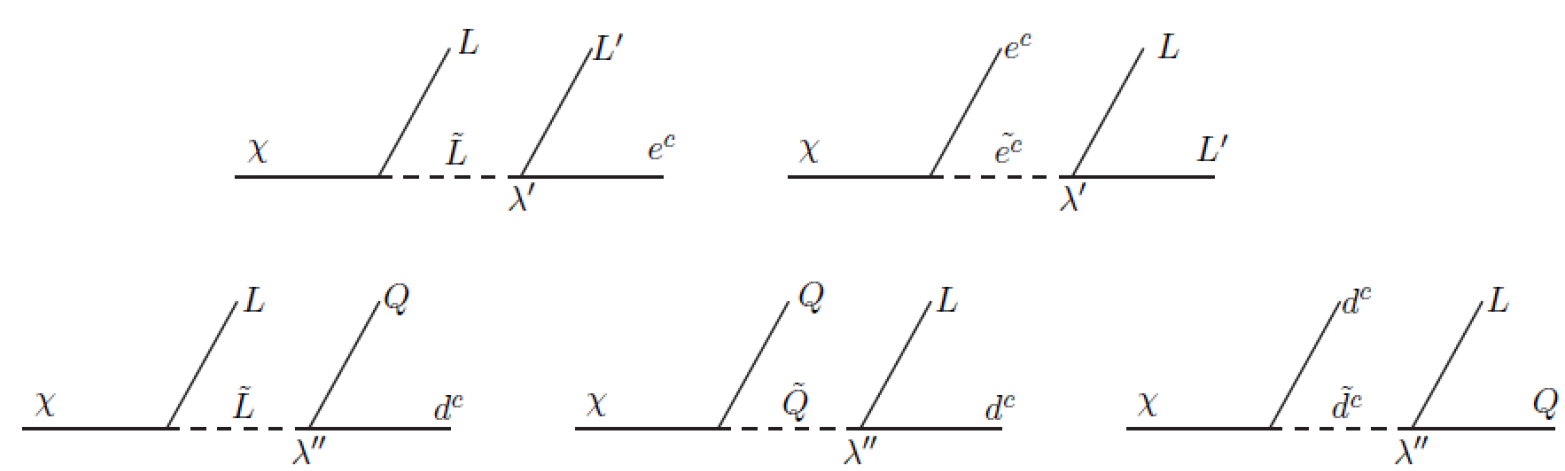}
\caption{Decays of the LSP. Only the lepton number violating diagrams are shown, since the lepton number violating couplings-- $\lambda^{\prime}$ (in the first line) and  $\lambda^{\prime\prime}$ (in the second line)-- receive large contributions (compared to the baryon number violating couplings) when the bilinear $R$-parity violating term, $h_u L$, is rotated away. Primes on the $L$ indicate that the lepton flavor is different than the slepton flavor. Figures from \cite{Martin:1997ns}.
\label{fig:rparityviolation}}
\end{figure} 

To be conservative in our estimates, we can take $\langle \nu^c \rangle \sim$ TeV, which may be expected from radiative symmetry breaking \cite{Ambroso:2010pe}. In this limit,   proton decay constraints are safe from $R$-parity violation, but there are more stringent constraints coming from the LSP lifetime. Current bounds on the LSP lifetime are slightly model dependent, but for the most part are \cite{Dreiner:1997uz}
\beq 
\tau_{LSP} \lesssim 1 \text{ sec} \;\;\;\;\;\;\mbox{OR}\;\;\;\;\;\; \tau_{LSP} \gtrsim 10^{25} \text{ sec }.
\eeq
The first bound excludes the region where the LSP decays would ruin the successful predictions of big bang nucleosynthesis on light nuclei abundances \cite{Reno:1987qw,Ellis:1990nb}. The other region is excluded by indirect dark matter detection experiments that search for energetic positrons and anti-protons coming from decaying or annihilating relics \cite{Berezinsky:1991sp,Baltz:1997ar,Arvanitaki:2009yb,Shirai:2009fq}.

The LSP lifetime can be calculated in terms of the general $R$-parity violating superpotential couplings (\ref{WRbreaking}). Diagrams in Figure 1 lead to an LSP lifetime
\beq
 \tau \approx  \frac{10^{-17} \sec}{ \lambda^{2}} \left(  \frac{m_{0}}{\mbox{ TeV}}    \right)^4  \left(  \frac {\mbox{100 GeV}}{ m_{\text{LSP}} }   \right)^5 
\eeq
where $\lambda  = \lambda^{\prime} ,\lambda^{\prime\prime}, \lambda^{\prime\prime\prime}$ and $m_0$ is the mass of the sfermion mediating the decay. Taking $\lambda =  \frac{\langle \nu^c \rangle}{m_p} \sim 10^{-15}$, $m_0 \sim 10$ TeV, and $m_{LSP} \sim 100 $ GeV gives 
\beq
\tau_{\text{LSP}} \sim 10^{17} \text{ sec },
\eeq
about the age of the universe. The $R$-parity violating couplings still need to be about $10^{-4}\sim 10^{-5} $ smaller to have an LSP lifetime greater than $10^{25}$ seconds. 

There are several ways additional suppressions might arise. We have not yet discussed the possibility of there being a horizontal family structure to the couplings. This could appear as a Froggett-Nielson symmetry, or a symmetry relating the locations of the matter singularities on the $G_2$ manifold, and would be responsible for forging the quark and lepton hierarchy. It may also suppress the LSP decay width pass the astrophysical bounds. Family symmetries  arise naturally from the $E_8$ structure \cite{King:2010mq}, which can also explain why the Standard Model has three generation, and this may hint towards a larger gauge theory. We leave this issue to future work..

If the family symmetry is not the answer, then it may be the case that resolution of the $E_8$ singularity to $SU(5)$ preserves a $U(1)$ symmetry--whose charges are necessarily given as a linear combination of four $U(1)$s belonging to the coset group $E_8/SU(5)$--and is broken to an exactly conserved $R$-parity. There are two well known examples,  $U(1)_\chi$ and $U(1)_\psi$ , defined as the symmetries coming from the breaking $SO(10)\rightarrow SU(5) \times U(1)_\chi$ and $E_6\rightarrow SO(10) \times U(1)_\psi$. However, $U(1)_\chi$ does not contain a field that can break $U(1)_\chi$ to $R$-Parity, and $U(1)_\psi$ forbids Higgs triplet masses, spoiling doublet-triplet splitting, so neither of these choices give a conserved $R$-parity.

However, there is a possibility that $U(1)$ symmetry is similar to $U(1)_\chi$, in that the MSSM fields and right handed handed neutrinos have the same charge assignment as in $U(1)_\chi$, but has additional $SU(5)$ singlet fields with even charges\footnote{If this $U(1)$ symmetry is to be broken to $R$-parity, then requiring the symmetry to be flavor blind, allowing for Higgs triplet masses,  and allowing an explanation for neutrino masses, basically constrains the charges of the MSSM and right handed neutrino fields to be the $U(1)_\chi$ charges.}.  These theories can then be broken to a {\it conserved} $R$-parity, when the additional singlets get vevs. It is easy to construct such a linear combination, though it is unclear why from a purely theoretical perspective why $G_2$ compactifications would favor this $U(1)$ symmetry. For instance, if $U(1)_a \times U(1)_b$ is the cartan subgroup of $SU(3)$ in the breaking pattern $E_8\rightarrow E_6\times SU(3)$, then the $U(1)$ given by the linear combination of charges
\[ q_\chi + 5(q_a-q_b) \]
allows for conical singularities that give rise to MSSM and right handed neutrino fields with $U(1)_\chi$ charges, but with additional $SU(5)$ singlets with charges $q=\pm 10$. The vevs of the additional singlets will break the $U(1)$ to a $Z_{10}$ symmetry that contains a $Z_2$ $R$-parity.

Finally we note (for the non string duality oriented reader) that $E_8 \times E_8$ is well motivated theoretically if the $G_2$-manifold is a $K3$ fibration. This is because the intersection matrix of 2-cycles inside $K3$ contain the Cartan matrix of $E_8 \times E_8$. It is in this case--that the gauge-theory of $M$ theory matches the gauge theory of $E_8 \times E_8$ Heterotic string theory-- in which  M theory on a $K3$-fibered $G2$-manifold and the heterotic string theory on a $T^3$-
fibered Calabi-Yau threefold are dual.

To summarize, we find that incorporating the $\mu$ parameter into the structure of $M$ theory compactified on a $G_2$-manifold, with stabilized moduli, can lead to a broken discrete symmetry allowing $\mu$ to be non-zero. $R$-parity is slightly broken, giving an LSP lifetime long enough to be the dark matter, but not quite long enough to evade satellite detector constraints. The theoretical structure allows for family symmetries, or an embedding of $R$-parity into $E_8$, both of which stabilize the LSP lifetime to be consistent with the experimental constraints. An example of the latter case is given above, so this is indeed a possibility. Either case will lead to the same dark matter phenomenology. The $R$-parity completion of this story is an interesting avenue for
further investigation.

\section{Phenomenology}

The $M$ theory framework, along with moduli stabilization in the $G_2$-MSSM, allows one to estimate the high-scale SUSY breaking masses and $\mu$  to within a factor of a few. This allows $M$ theory to make many phenomenological predictions. For some cases even small variations in the high-scale theory can have significant phenomenological consequences. In particular, the low-scale values of $\mu $ and $\tan\!\beta$ have significant implications for dark matter properties, and thus it is crucial to have a good understanding of their low-scale values while considering the $M$ theory predictions of the high-scale masses.

\subsection{Electroweak Symmetry Breaking}

The first and foremost phenomenological constraint is that the theory accurately produce electroweak symmetry breaking (EWSB). That is, the theory must give a stable potential (bounded from below), break the electroweak symmetry and allow for the correct Z-boson mass. Respectively, these three conditions can be quantified by the following tree level constraints at the EWSB scale
\beq\begin{array}{ccc}
\vspace{3 mm}
 \left| B\!\mu \right| & \le &  \frac{1}{2}( m_{H_u}^2 + m_{H_d}^2 ) + |\mu |^2 \\ 
\vspace{1 mm}
 \left| B\!\mu \right|^2  & \ge & ( m_{H_u}^2 +  |\mu |^2)( m_{H_d}^2 + |\mu |^2) \\
M_Z^2 & = &  -2  |\mu |^2 + 2 \dfrac{m_{H_d}^2 -  m_{H_u}^2 \tan^2\beta}{ \tan^2\beta-1}
\label{EWSBconstraints}
\end{array}\eeq
where $\tan\!\beta$ is not an independent parameter, but is determined by
\beq
\sin 2\beta = \frac{-2 B\!\mu}{m_A^2 } . \label{bmutanbeta}
\eeq
and 
\beq
 m_A^2 = m_{H_u}^2 + m_{H_d}^2  +2 |\mu |^2 \label{mA}.
\eeq 
where $A$ is the pseudoscalar Higgs boson.

To get a feeling for $\tan\!\beta$, we plug in the expected values (at the unification scale and with degenerate scalars) of $B\!\mu \simeq 0.2 m_{3/2}^2$ and $m_A^2 \simeq 2 m_{3/2}^2$, into ($\ref{bmutanbeta}$) which gives  $\tan\!\beta \simeq 10$.   On the other hand, RGE flow will lower the values of both $B\!\mu $ and $m_A^2$, resulting in variations around $\tan\!\beta \simeq 10$. In Section $7.1.1$, a numerical scan will show a lower bound of $\tan\!\beta \gtrsim 5$, when scalars are taken to be degenerate at the unification scale.  

The lowest values of $\tan\!\beta$ occur for the smallest values of  $m_A^2$. The EW scale value for the mass depends on the running of the Higgs scalar masses, and in turn is very sensitive to the values of the squark masses. For specific non-degenerate values of scalar masses at the unification scale,  $m_A^2$ can be of order $B\!\mu$ at the EW scale, resulting in values of $\tan\!\beta < 5$. We will consider this situation in Section $7.1.2$.

At tree level the mass of the Z-boson is determined by the four Higgs  parameters
\begin{equation}   M_Z (m_{H_u}^2, \; m_{H_d}^2, \;   |\mu |^2,  \; \tan\!\beta). \end{equation}
These parameters not only depend on their respective values at the high-scale, but also on other masses as a result of RGE-flow. Assuming that the scalar masses are much larger than the gaugino masses, $M_Z$ has strongest dependence on the Higgs mass parameters and stop masses
\beq  M_Z( \hat{m}_{H_u}^2, \; \hat{m}_{H_d}^2, \;  \hat{B\mu}, \; |\hat{\mu} |^2, \; \hat{m}_{Q_3}^2, \; \hat{m}_{U_3}^2 ) .\eeq
where hatted ($~\hat{}~$) masses refer to GUT scale values. 

Interestingly the cancellation between the soft scalars masses contributing to $M_Z$ can be significant, even in the case in which the scalar masses are unified at the GUT scale
\[ \hat{m}_{H_u}^2 = \hat{m}_{H_d}^2 = \hat{m}_{Q_3}^2 = \hat{m}_{U_3}^2.\]
Naively what one thought was a large fine-tuning between the Higgs soft-masses and $\mu$ in eq. (\ref{EWSBconstraints}) for $M_Z$, is in fact smaller. This is evident (see Figure \ref{fig:unifiedmutanbeta})  from the fact that the scalar masses can be of order the gravitno mass at unification and $\mu$ can be an order of magnitude smaller, but the cancellation in eq. (\ref{EWSBconstraints}) for $M_Z$ still occurs. In this sense, the ratio $\mu / m_{3/2}$, shown in Figure \ref{fig:unifiedmutanbeta}, might be considered a measure of the fine-tuning involved in EWSB. In other words, the smaller the ratio, the less fine tuning there will be of $\mu$ against the scalar masses in order to have the correct value for $M_Z$.

\subsubsection{Degenerate Scalars}
A numerical scan was performed over $M$ theory parameter space described in \cite{Acharya:2008zi} using SOFTSUSY \cite{Allanach:2001kg}\footnote{See \cite{Feldman:2010uv} for general phenomenological discussions.}. We allow for the following variation in the  $G_2$-MSSM parameters, 
\begin{itemize}
\item $10 \tev \le  m_{3/2} \le  20 \tev $ -- the gravitino mass
\item $10\le V_7 \le 40$ -- the volume of the $G_2$-manifold in units of the eleven-dimensional Planck length
\item  $-10 \le \delta \le 0$ -- the size of the threshold corrections to the (unified) gauge coupling, $\alpha^{-1}_{\text{GUT}}$. \footnote{see Section IV  of  \cite{Acharya:2008zi} for the precise definition of $\delta$}
\end{itemize}
An interested reader is referred to Section V of  \cite{Acharya:2008zi} for variations in the spectra of $G_2$-MSSM models. In addition, order one variations are allowed for the coefficients in (\ref{muterm}) for the formula for $\mu$,  while it is imposed that $B\!\mu$ is in the range.
\beq 
1\,\mu m_{3/2} < B\!\mu < 3\,\mu m_{3/2}. \label{bmubound}
\eeq

\begin{figure}[htp]
\centering
\includegraphics[width=1\textwidth]{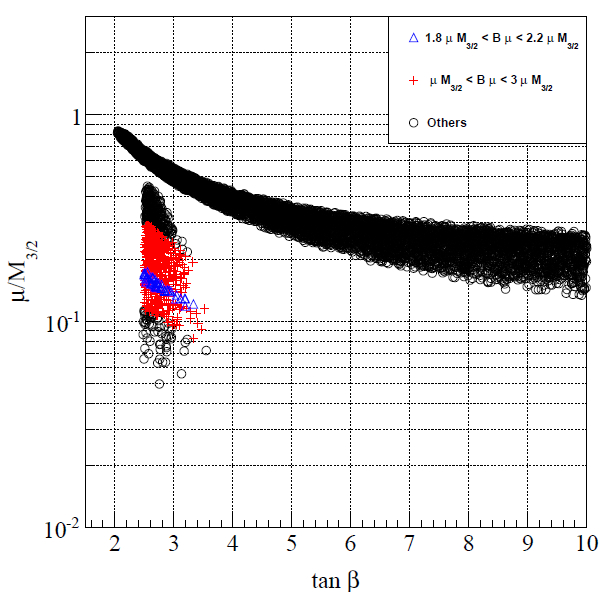}
\caption{$\mu/m_{3/2}$ vs. $\tan\!\beta$. The upper band scans over the $G_2$-MSSM parameter space with {\emph {degenerate scalars}} at the unification scale. The lower region on the left (low $\tan\!\beta$)  scans over the $G_2$-MSSM parameter space where the scalar mass ratio  $\hat{m}^2_{H_u}\!:\! \hat{m}^2_{U_3} \!:\! \hat{m}^2_{Q_3} = 3\!:\! 2\!:\! 1$ is required to be accurate within $20\%$. The black points show models that correctly break the EW symmetry, but are  inconsistent with constraint $1\,\mu m_{3/2} < B\!\mu < 3\,\mu m_{3/2}$, so we expect them to not be valid solutions. The red points satisfy the constraint on $B\!\mu$ as given in the legend.  The empty space on the plot, between the two regions, is expected to be filled in with complete scan over possible non-degenerate scalar mass parameter space. All points have EWSB.}
\label{fig:unifiedmutanbeta}
\end{figure}

The results are shown in Figure~\ref{fig:unifiedmutanbeta}. As is evident from the plot, values of $\mu$ much smaller than the gravitino mass are allowed under all the constraints, signaling a non-imposed  cancellation among the scalars  contributing to $M_Z$. 
Of note is the fact that $\tan\!\beta$ and $\mu$ are inversely correlated, which will play a significant role in limiting the maximum spin-independent scattering cross-section, when scalar masses are unified at the high scale.

\subsubsection{Non-Degenerate Scalars and Low {\boldmath{$\tan\!\beta$}}}

We also consider the possibility that $M$ theory allows for scalar unification to be somewhat perturbed (at the factor of two to three level). Since eventually we will be interested in calculating the largest possible spin-independent scattering cross sections we will only consider high-scale scalar masses that give $\tan\!\beta\lesssim 3 $--since the scattering cross sections decrease with increasing $\tan\!\beta$

Consider the 1-Loop RGE equations, where only terms proportional to $\lambda_t$ are kept, while neglecting the $\lambda_t$ running. The RGE equations of the relevant scalars are:
\beq \begin{array}{ccl}
    {8\pi^2}\frac{d m_{H_u}^2}{dt} &= &3 \left| \lambda_t \right|^2 \left( m_{H_u}^2 + m_{Q_3}^2 + m_{U_3}^2 + \left| A_t \right|^2\right) \nonumber \\
    {8\pi^2}\frac{d m_{{U_3}}^2}{dt} &= &2 \left| \lambda_t \right|^2 \left( m_{H_2}^2 + m_{Q_3}^2 + m_{U_3}^2 + \left| A_t \right|^2\right) \nonumber \\
   {8\pi^2}\frac{d m_{{Q_3}}^2}{dt} &=& 1 \left| \lambda_t \right|^2 \left( m_{H_2}^2 + m_{Q_3}^2 + m_{U_3}^2 + \left| A_t \right|^2\right) \nonumber \\
    {8\pi^2}\frac{dA_t}{dt} &=& 6\lambda_t^2 A_t
    \label{eqn:rge}
\end{array}\eeq
whose solution is 
\beq\begin{array}{ccl}
m^2_{H_u} &=& \frac{1}{2} \left( \hat{m}^2_{H_u} -\hat{m}^2_{U_3} -  \hat{m}^2_{Q_3}  + e^{\frac{3 t \lambda^2}{4 \pi^2}}( |\hat{A}_t|^2 (-1+ e^{\frac{3 t \lambda^2}{4 \pi^2}}) +  \hat{m}^2_{H_u} +  \hat{m}^2_{U_3}+  \hat{m}^2_{Q_3}) \right) \\
m^2_{U_3} &=& \frac{1}{3} \left(- \hat{m}^2_{H_u}+ 2\hat{m}^2_{U_3} -  \hat{m}^2_{Q_3}  + e^{\frac{3 t \lambda^2}{4 \pi^2}}(|\hat{A}_t|^2 (-1+ e^{\frac{3 t \lambda^2}{4 \pi^2}}) + \hat{m}^2_{H_u} +  \hat{m}^2_{U_3}+  \hat{m}^2_{Q_3}) \right) \\
m^2_{Q_3} &=& \frac{1}{2} \left(- \hat{m}^2_{H_u}  - \hat{m}^2_{U_3} + 5 \hat{m}^2_{Q_3} + e^{\frac{3 t \lambda^2}{4 \pi^2}}(|\hat{A}_t|^2 (-1+ e^{\frac{3 t \lambda^2}{4 \pi^2}}) + \hat{m}^2_{H_u} +  \hat{m}^2_{U_3}+  \hat{m}^2_{Q_3} ) \right) \\
A_t^2 &=& \hat{A}_t^2 e^{\frac{3 t \lambda^2}{8 \pi^2}} \\
\label{rgesolutions}
\end{array}\eeq
where hatted ($\hat{}$) masses indicate GUT scale mass.

Since $m^2_{H_d}$ barely runs for low $\tan\!\beta$ and it is predicted that $\hat{\mu}^2$ is over an order of magnitude smaller than  $m^2_{H_d}$ , the cancellation in $M_Z$ (\ref{EWSBconstraints}) should occur between  $m^2_{H_u}$ and $m^2_{H_d}$. Therefore,  $m^2_{H_u}$ needs to stay positive at the EWSB scale. Ignoring the exponentially suppressed terms in (\ref{rgesolutions}), we see that there are no choices of $\{ \hat{m}^2_{H_u}, \hat{m}^2_{Q_3},\hat{m}^2_{U_3} \} $ that leave all low-scale masses positive. On the other hand, there is a fixed point solution to the above RGEs
\beq
 \hat{m}^2_{H_u}: \hat{m}^2_{U_3} : \hat{m}^2_{Q_3} = 3 : 2: 1 \label{fixedpoint}
\eeq
where the non-exponentially suppressed terms are identically zero, insuring that if the trilinears are of order the scalars as expected in the $G_2$-MSSM, all three masses will stay positive. This fixed point is analogous to the focus point solution in minimal supergravity (mSUGRA) theories \cite{Feng:1999zg,Chan:1997bi}, as it minimizes the fine tuning of EWSB. However, unlike the focus point region of mSUGRA where the Higgs scalars run small due to RGE flow, here the scalars remain heavy, and are close to the gravitino mass. 

 Near this region, low $\tan\!\beta$ parameter space with EWSB can be realized. Results on the numerical scan can be seen in Figure~\ref{fig:unifiedmutanbeta}.

\subsection{The Nature of the LSP}
As explained in detail in\cite{Acharya:2008zi}, the $G_2$-MSSM framework gives
rise to mostly Wino LSPs (as opposed to Bino LSPs). The tree
level gaugino masses are degenerate at the GUT scale, but are suppressed by $F$-terms of the moduli relative to the gravitino mass to be of order the gaugino masses from the anomaly mediation contribution. The additional contribution from the anomaly lifts $M_1$ over $M_2$, leading to mostly Wino LSP models. In the original  $G_2$-MSSM scenario, where it was simply that $\mu \sim m_{3/2}$, there were additional contributions to the gaugino masses from supersymmetric Higgs loops, proportional to $\mu$ \cite{Pierce:1996zz}, that for some choices of high scale parameters, could re-lift $M_2$ over $M_1$. These models are disfavored by precision gauge coupling unification \cite{Acharya:2008zi}, and occur less frequently in parameter space here than the original models since $\mu \not{\!\!\!\sim}\, m_{3/2}$. However, smaller $\mu$ will tend to introduce a small Higgsino admixture into the mostly Wino LSP - a fact which has significant implications on dark matter discovery (Section 6.3). All these considerations combine to strongly suggest that a Wino-like LSP with mass $\sim 140 - 200$ GeV constitutes a significant fraction of the dark matter.

As emphasized in \cite{Acharya:2008bk,Acharya:2009zt}, in order to obtain about the right relic density from the moduli decays,
the LSP must be a Wino-like particle, with a large
annihilation cross section of about $3 \times 10^{−24} \cm^2$. A non-thermal
history dominated by moduli and a wino LSP give a consistent picture
for dark matter from the compactified string theory. Also
encouraging is the fact that the PAMELA satellite data on
positrons and antiprotons can
be consistently described by a Wino LSP \cite{Grajek:2008jb,Hisano:2008ti,Kane:2009if,Feldman:2009wv,Chen:2010yi,Chen:2010kq}.
More recently, by also considering Wino annihilations into photons and Z-bosons one finds a cross-section of
about $10^{-26} \cm^2$ -- a fact relevant for future Fermi data.

\subsection{Direct Detection of Dark Matter}

In December 2009, CDMS reported at most two possible WIMP candidate events, with a high likelihood of being background \cite{Ahmed:2009zw}. Combining  with their previous data, this amounts to a bound on the spin-independent scattering cross-section of $\sigma_{si} \lesssim 6 \times 10^{-44} \text{ cm}^2 $ for a WIMP of mass around $200$ GeV. More recently, the XEXON100 experiment \cite{Aprile:2010um} reported observing no events after their first 11 days of running, slightly strengthening the CDMS bound. In the near future, XEXON100 is expected to report results that will probe much smaller scattering cross sections $\sigma_{SI} \sim 2\times 10^{-45}  \text{cm}^2 $. We will see that even this region is out of reach given the $M$ theory predictions we calculate.

In the decoupling limit, defined when the pseudoscalar mass is much larger that the Z-boson mass, $m_{A^0} \gg M_{Z}$, the charged and heavy CP-even Higgses are also heavy, $m_{H^{\pm}} \simeq m_{H^{0}}  \simeq m_{A^{0}} $. The other Higgs boson $h^0$ remains light and behaves in the same way as the SM Higgs boson. The lower bound on its mass, corresponds to the same bound on the SM Higgs boson, namely 114 GeV\footnote{Since there are theoretical and calculational uncertainties with calculating the Higgs mass, we will consider models with $m_h \ge 110$ GeV.}\cite{Barate:2003sz}. All the models consistent with all the theoretical and phenomenological constraints have light Higgs mass close to thia LEP limit. Since the squarks are also heavy in $G_2$-MSSM, the light Higgs boson exchange will give the only substantial contribution to the spin-independent scattering cross sections. The scattering of the LSP off nuclei is via the Higgsino component. While the LSP will be mostly Wino-like, the prediction that  $\mu$ is of order the TeV scale implies that the LSP wavefunction can have non-trivial Higgsino mixing.

Following \cite{Cohen:2010gj} we estimate the size of the direct detection cross section in the decoupling limit to be
\begin{equation}
   \sigma_{\rm SI} \left( \chi N \rightarrow \chi N \right) \approx 5 \times 10^{-45} \text{cm}^2 \left( \frac{115\gev}{m_h} \right)^4
 \left(\frac{Z_{H_u}\sin\beta - Z_{H_d}\cos\beta}{0.1}\right)^2 \left( Z_W - \tan\theta_W Z_B \right)^2
    \label{eqn:tim}
\end{equation}
where the $Z$'s give the composition of the LSP
\beq
\chi \equiv Z_B\,\tilde{B}+Z_W\,\tilde{W}\,+Z_{H_d}\,\tilde{H}_d+Z_{H_u}\,\tilde{H}_u.
\eeq

This gives us an estimate of the largest direct detection scattering cross sections, which naively may seem that for $Z_{H_u} \sim 0.1$ can be very close to the reach of XENON. Eq. (\ref{eqn:tim}) can further be simplified, with the aid of analytical expressions for the neutralino mass matrix eigenvalues and eigenvectors \cite{ElKheishen:1992yv,Barger:1993gh,Bertone:2004pz}. Taking the limit $ M_1 = M_2 $, which maximizes the scattering cross section for fixed $\mu$ and $\tan\!\beta$, (\ref{eqn:tim}) becomes 
\begin{equation}
    \sigma^{\rm MSSM}_{\rm SI} \left( \chi N \rightarrow \chi N \right) \approx 6 \times 10^{-45} \text{cm}^2 \left( \frac{115 \gev}{m_h} \right)^4    
\left(\frac{1 \tev}{\mu}\right)^2 \left( \frac{\sin2\beta+M_{2}/\mu}{1 - (M_2/\mu)^2} \right)^2
    \label{eqn:upperlimit}
\end{equation}
which falls off both with $\tan\!\beta$ and $\mu$. Allowing for the variation in $M_1$ and $M_2$ in the $G_2$-MSSM will only decrease this fraction. The value $M_2/\mu$ is typically around $.1\sim.2$. The parameters for three different models, along with their scattering cross sections, can be seen in Table 1 and are appropriately labeled in Figure 3.
 
\begin{figure}[hpt!]
\centering
\includegraphics[width=1\textwidth]{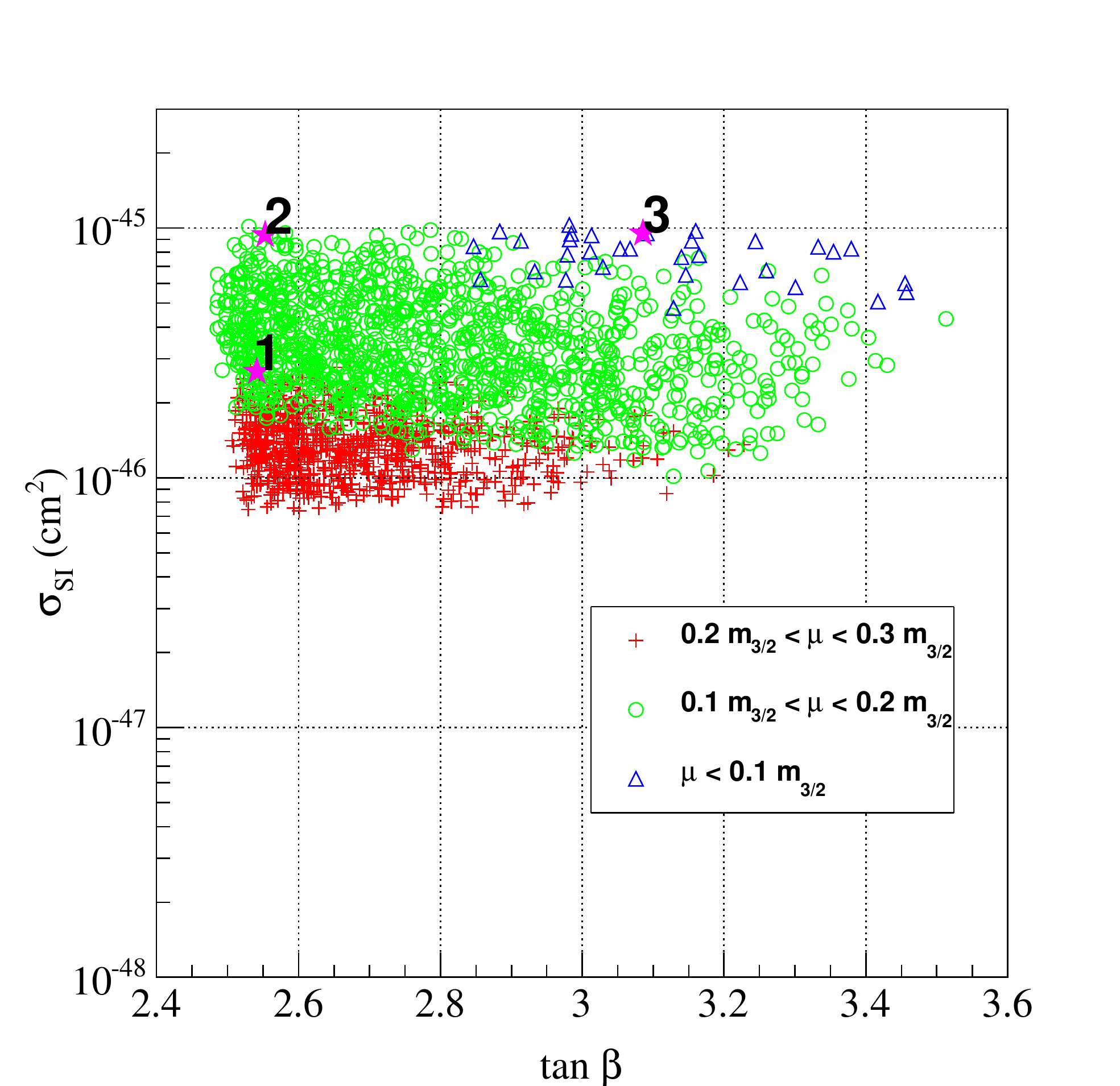}
\caption{Spin-independent scattering cross-sections vs  $\tan\!\beta$. The region shown  scans over the $G_2$-MSSM parameter space where the scalar mass ratio  $\hat{m}^2_{H_u}\!:\! \hat{m}^2_{U_3} \!:\! \hat{m}^2_{Q_3} = 3\!:\! 2\!:\! 1$ is required to be accurate within $20\%$. All points satisfy the constraint $\mu m_{3/2} < B\!\mu < 3\,\mu m_{3/2}$, have a SM-like Higgs with mass $m_h \ge 110$ GeV, and have EWSB.  We also list the parameters for the 3 models in Table 1. In the region where EWSB, supergravity, and phenomenological constraints are satisfied, the upper limit on $\sigma_{SI}$ is robust, but the lower limit can decrease if the sign of $\mu$ is reversed.}
\label{fig:xsecuniversal}
\end{figure}

However, as shown in the previous section, when considering degenerate scalar masses at the unification scale EWSB imposes that small $\mu$ corresponds to large $\tan\!\beta$, and small $\tan\!\beta$ corresponds to large $\mu$. Hence, large cross-sections, of order the XEXON100 reach are not attainable for this region. To verify this we perform a scan of parameter space, using DarkSUSY \cite{Gondolo:2004sc}. The results are show in Figure \ref{fig:xsecuniversal} where it is seen that the largest scattering cross-sections are $\sim 1 \times 10^{-45} \text{cm}^2$, close to, but slightly beyond the reach of XENON100.

In Figure~\ref{fig:xsecuniversal} we also scan over  the $G_2$-MSSM parameter space, while requiring that the ratio $ \hat{m}^2_{H_u}: \hat{m}^2_{U_3} : \hat{m}^2_{Q_3} = 3 : 2: 1$ be accurate within $20\%$. The spin-independent scattering cross-section reaches an upper limit of $1 \times 10^{-45}\cm^2$, just beyond the XENON100 reach. Since this is the region where largest cross-sections appear, we can conclude that if the solution of the $\mu$-problem proposed, along with moduli-stabilization in the $G_2$-MSSM, is the model of nature, the XENON100 experiment will not observe a dark matter signal soon, but its next run and upgraded detectors may do so.

\begin{table}
\setlength{\columnsep}{145pt}
\begin{multicols}{2}

\begin{tabular}{|c|ccc|}
\hline
   & Model 1 & Model 2  & Model 3  \\
\hline
$M_{\text{3/2}}$ 			& 17.8 TeV	&18.1 TeV	&17.9 TeV		 \\
$\sqrt{B\mu_{\text{GUT}}}$ 	& 9.75 TeV	&10.4 TeV	&9.29 TeV		 \\
$ \mu_{\text{GUT}}$ 		& 3.79 TeV	& 2.10 TeV	&1.69 TeV		 \\
\hline
$ M_1 $  				& 151. GeV	& 153. GeV	&150. GeV		 \\
$ M_2 $  				& 145. GeV	& 143. GeV	&138. GeV		 \\
$ \mu$  				& 3.89 TeV	& 2.15 TeV	&1.77 TeV		 \\
$ M_A $  				& 18.8 Tev	& 19.0 TeV	&18.2 TeV		 \\
$ m_h $  				& 110. GeV	& 110. GeV	&115. GeV		 \\
\hline
$ M_{\chi_1 } $  			& 141 GeV	& 143 GeV	&141 GeV		 \\
$ M_{\chi_2 } $  			& 143 GeV	& 147 GeV	&145 GeV		 \\
$ M_{\chi_1^\pm } $  		& 141 GeV	& 144 GeV	&142 GeV		 \\
\hline
$ Z_{\tilde{W}}$  			& 0.94	& 0.91	&0.91			 \\
$ Z_{\tilde{B}} $  			& -0.35	& -0.41	&-0.41		 \\
$ Z_{\tilde{H}_d} $  		& -0.02	& -0.04	&-0.05			 \\
$ Z_{\tilde{H}_u} $  		& 0.01	& 0.02	&0.02		 \\
\hline
$ \tan\beta $  	 		& 2.53	& 2.37	&2.87			 \\
$ \sigma _{\text{SI}} [\text{cm}^2]$	 	&$3.\times 10^{-46}$&$ 9.\times 10^{-46}$&$ 1.\times 10^{-45}$  \\
$ \sigma _{\text{SD}} [\text{cm}^2]$	 	&$5.\times 10^{-45}$&$ 5.\times 10^{-44}$&$ 1.\times 10^{-43}$  \\
\hline
\end{tabular}
\\
~\\
 \caption{ High scale and low scale parameters for 3 models with larger spin independent scattering cross sections. All models shown belong to the parameter space where the scalar mass ratio  $\hat{m}^2_{H_u}\!:\! \hat{m}^2_{U_3} \!:\! \hat{m}^2_{Q_3} = 3\!:\! 2\!:\! 1$  is accurate within $20\%$. We assume that details of the calculation and software outputs are sufficiently uncertain to allow $m_h \gtrsim 110$ GeV to be consistent with LEP bounds. }
\end{multicols}
\end{table}

\newpage

\section{Conclusions}

We have argued that if our universe is described by $M$ theory compactified
on a manifold of $G_2$ holonomy, with doublet-triplet splitting solved in the way originally proposed by Witten \cite{Witten:2001bf}, 
then there is a simple solution to the $\mu$-problem:
strong coupling dynamics in the the hidden sector will generate a non-perturbative potential for the moduli, which stabilizes all the moduli vevs, and breaks the symmetry forbidding $\mu$.
Then, following the numerical analysis done in the $G_2$-MSSM \cite{Acharya:2008zi}, the breaking will generate $\mu \sim \langle \frac{S}{m_{pl}} \rangle m_{3/2} \sim 0.1 ~m_{3/2} \sim 2 \tev$. 
This then implies a non-zero Higgsino component of the mostly Wino LSP, with an upper limit, which in turn gives an upper limit of about $1 \times 10^{-45}\cm^2$ on the spin-independant scattering cross-section, somewhat below the reach of the XENON100 experiment, as well as a lower limit of about $10^{-46} \cm^2$. The Wino-like LSP also can account for the PAMELA positron and antiproton excesses \cite{Adriani:2008zr,Kane:2009if}, and gives about the desired relic density for a non-thermal cosmological history \cite{Acharya:2010af}, as expected in theories with moduli.

Since the scalars are of order $m_{3/2} \gtrsim 20 $ TeV, the Higgs sector is decoupled, and the light Higgs boson behaves like a Standard Model one. It's mass is predicted to be of order 110-120 GeV. If we insist on a good description of the Pamela data plus consistent compactification, we find an LSP mass from about 140-155 GeV, and an annihilation cross section $2-3.5 \times 10^{-24} \text{cm}^3/\text{s}$.  The annihilation to $\gamma/Z$ ranges from $(0.7-1.2) \times 10^{-26}$. 

Additionally, we noted that an eact $R$-parity could arise through `partial symmetry breaking', though this isn't obviously motivated by the theory itself. An alternative is that $R$-parity is either an exact remnant of a broken continuous gauge symmetry, or only an approximate symmetry of larger broken discrete and continuous groups. In either case, this requires the inclusion of additional $U(1)$ gauge symmetries, suggesting that the GUT group is larger than $SU(5)$, and may originate from an $E_8$ singularity.

\section*{Acknowledgments}

We appreciate helpful conversations with Tim Cohen, Daniel Feldman, Piyush Kumar, Paul Langacker, Joseph Marsano, Aaron Pierce, and Lian-Tao Wang. B.A..
is grateful to the University of Michigan Physics Department and MCTP for support, and E.K. is grateful for a  String Vacuum Project Graduate Fellowship
funded through NSF grant PHY/0917807. This work was supported by the DOE Grant \#DE-FG02-95ER40899.

\section*{Appendix A: Largest Spin Independent Cross Sections}

Following \cite{Cohen:2010gj},  the spin-independent cross section for the LSP scattering off a nucleon, is given in the decoupling limit ($M_Z \ll M_A$) by the approximation
\begin{equation}
   \sigma_{\rm SI} \left( \chi N \rightarrow \chi N \right) \approx 5 \times 10^{-45} \text{cm}^2 \left( \frac{115\gev}{m_h} \right)^4
 \left(\frac{Z_{H_u}\sin\beta - Z_{H_d}\cos\beta}{0.1}\right)^2 \left( Z_W - \tan\theta_W Z_B \right)^2
    \label{eqn:tim}
\end{equation}
where the $Z$'s give the composition of the LSP
\begin{equation}
\chi \equiv Z_B\,\tilde{B}+Z_W\,\tilde{W}\,+Z_{H_d}\,\tilde{H}_d+Z_{H_u}\,\tilde{H}_u.
\end{equation}

Consider the neutralino mass matrix \cite{Frere:1983dd}:
\begin{equation}
    \mathcal{M} =
    \begin{pmatrix}
        M_1 & 0 & -M_Z \cos\beta\sin\theta_W & M_Z\sin\beta\sin\theta_W \\
        0 & M_2 & M_Z \cos\beta\cos\theta_W & -M_Z\sin\beta\cos\theta_W \\
        -M_Z\cos\beta\sin\theta_W &  M_Z\cos\beta\cos\theta_W & 0 & -\mu \\
        M_Z\sin\beta\sin\theta_W & -M_Z\sin\beta\cos\theta_W & -\mu & 0
    \end{pmatrix}
    \label{eqn:massmatrix}
\end{equation}
in the $\{\tilde{B},\tilde{W},\tilde{H}_d,\tilde{H}_u\}$ basis.
The analytical expression \cite{ElKheishen:1992yv,Barger:1993gh,Bertone:2004pz} for the components in the LSP can be written as:
\begin{align}
    \alpha Z_{B} &= z_B =  - \sin\theta_W \nonumber\\
     \alpha Z_{W} &= z_W = \cos\theta_W \frac{M_1-M_\chi}{M_2-M_\chi} = \cos\theta_W \frac{\left( M_1- M_\chi \right)^2}{\Delta} \nonumber \\
     \alpha Z_{H_d} &= z_ {H_d}  = \frac{\mu\left( M_1 - M_\chi \right)\left( M_2 - M_\chi \right) + M_Z^2\sin\beta\cos\beta\left( \left( M_1 - M_2 \right)\cos^2\theta_W +M_2 - M_\chi \right)}{M_Z\left( M_2 - M_\chi  \right)\left( -\mu\cos\beta + M_\chi\sin\beta \right)} \nonumber\\
     \alpha Z_{H_u} &= z_{H_u} = \frac{M_\chi \left( M_1 - M_\chi \right)\left( M_2 - M_\chi \right) + M_Z^2\cos^2\beta\left( \left( M_1 - M_2 \right)\cos^2\theta_W +M_2 - M_\chi \right)}{M_Z\left( M_2 - M_\chi  \right)\left( -\mu\cos\beta + M_\chi\sin\beta \right)}
    \label{eqn:components}
\end{align}
where $ \alpha = \sqrt{ z_{B}^2+z_{W}^2+z_{H_d}^2+ z_{H_u}^2}$ is an overall normalization factor and $\Delta\equiv(M_\chi -M_1)(M_\chi-M_2)$. 

The combination $Z_{H_u}\sin\beta-Z_{H_d}\cos\beta$, that appears in the scattering cross section takes an especially simple form
\begin{equation}
    Z_{H_u}\sin\beta-Z_{H_d}\cos\beta = \frac{\left( M_\chi\sin\beta - \mu\cos\beta \right)\left( M_1-M_\chi \right)\left( M_2-M_\chi \right)}{M_Z\left( M_2 - M_\chi \right)\left( -\mu\cos\beta+M_\chi\sin\beta \right)} = \frac{M_1-M_\chi}{M_Z}.
    \label{eqn:components2}
\end{equation}
It is clear from (\ref{eqn:components2}) that as $M_1 - M_\chi$ increases, $Z_{H_u}\sin\beta-Z_{H_d}\cos\beta$ grows slower than the $Z_{W}$ component. Thus after normalization both the $\tilde{H}_u$ and the $\tilde{H}_d$ components will decrease. So the maximum of $Z_{H_u}\sin\beta-Z_{H_d}\cos\beta$ is realized when $M_1 - M_\chi$ is minimal.

The eigenvalues of the neutralino mass matrix (\ref{eqn:massmatrix}) are given by the solutions to:
\begin{equation}
    \left( x - M_1 \right)\left( x - M_2 \right)\left( x - \mu \right)\left( x + \mu \right)+\left(M_1\cos^2\theta_W +M_2\sin^2\theta\right)M_Z^2\mu\sin2\beta = 0
    \label{eqn:eigenvalue}
\end{equation}
Then the LSP mass, corresponding to $M_{\chi} \equiv x$, can be found by taking the limit $M_\chi \ll \mu$, so that (\ref{eqn:eigenvalue}) is simply a quadratic equation. Then it is easy to see that the minimal value of $M_1 - M_\chi$, which maximizes $Z_{H_u}\sin\beta-Z_{H_d}\cos\beta$, corresponds to the situation when $M_1 - M_2$ is also minimized. Additionally, when $M_1 = M_2$, the term $Z_W - \tan\theta_W Z_B$ also reaches its maximum. Thus the maximum scattering cross sections will occur when $M_1 = M_2$.

To normalize the expressions in (\ref{eqn:components}) (i.e. finding $\alpha$) is tedious. Instead, a new basis is defined where $\tilde{\gamma} = \cos\theta_W \tilde{B} + \sin\theta_W \tilde{W}$ and $\tilde{Z} = -\sin\theta_W \tilde{B} + \cos\theta_W \tilde{W}$, where in the supersymmetric limit, these are the superpartners of the photon and $Z$-boson, respectively. The new mass matrix, in the $\{ \tilde{\gamma},\tilde{Z},\tilde{H}_d,\tilde{H}_u \}$  is 
\begin{equation}
    \mathcal{M} =
    \begin{pmatrix}
        M_1{\cos\theta_W}^2 +M_2{\sin\theta_W}^2 & (M_2-M_1)\sin\theta_W\cos\theta_W & 0 & 0 \\
        (M_2-M_1)\sin\theta_W\cos\theta_W & M_2{\cos\theta_W}^2 +M_1{\sin\theta_W}^2  & M_Z \cos\beta  & -M_Z\sin\beta \\
       0 &  M_Z\cos\beta & 0 & -\mu \\
       0 & -M_Z\sin\beta & -\mu & 0
    \end{pmatrix}
    \label{eqn:massmatrix2}
\end{equation}
Taking the limit $M_1 = M_2 \equiv M$, one immediately one finds that $\tilde{\gamma}$ in an eigenvector with mass eigenvalue $M$. The next lightest eigenvector of the remaining $3\times 3$ sub-matrix will be mostly $\tilde{Z}$, and to leading order in $M_Z/\mu$, the mass is
\beq
 M_\chi \simeq M - \frac{M_Z^2}{\mu}\left( \frac{M}{M_Z}-\sin 2 \beta \right)
\eeq
Next we will assume that the phases of $M$ and $\mu$ are such that absolute value of $ M_\chi$ is smaller than $|M|$, so that it is indeed the LSP.  The other scenario, in which the LSP is mostly $\tilde{\gamma}$, will have negligible scattering cross-section.

Diagonalizing the remaining $3\times 3$ sub-matrix, the coefficients of $\tilde{H}_u$ and $\tilde{H}_d$ component to leading order is
\begin{align}
    Z_{H_d} &= \frac{1}{2}\left( \frac{\left(\sin\beta - \cos\beta\right)M_Z}{\mu+M} + \frac{\left(\sin\beta + \cos\beta\right)M_Z}{\mu - M} \right) = \frac{M_Z\left( \mu\sin\beta+M\cos\beta \right)}{\mu^2 - M^2} \nonumber \\
    Z_{H_u} &= \frac{1}{2}\left( \frac{\left(\sin\beta - \cos\beta\right)M_Z}{\mu+M} -\frac{\left(\sin\beta + \cos\beta\right)M_Z}{\mu - M} \right) = -\frac{M_Z\left( \mu\cos\beta+M\sin\beta \right)}{\mu^2 - M^2}
    \label{eqn:hu}
\end{align}
and from the definition on $\tilde{Z}$,
\beq Z_W - \tan\theta_W Z_B = {\cos\theta_W}^{-1}.     \label{eqn:zu} \eeq

Finally, using (\ref{eqn:hu}) and (\ref{eqn:zu}) as inputs to (\ref{eqn:tim}) the upper limit for the cross section is
\begin{equation}
    \sigma_{\rm SI} \left( \chi N \rightarrow \chi N \right) \approx 6 \times 10^{-45} \text{cm}^2 \left( \frac{115 \gev}{m_h} \right)^4    
\left(\frac{1 \tev}{\mu}\right)^2 \left( \frac{\sin2\beta+M_{2}/\mu}{1 - (M_2/\mu)^2} \right)^2
    \label{eqn:upperlimit}
\end{equation}
From the discussion in the text we expect $M_2 /\mu \lesssim 0.2$, $\sin{2\beta} \lesssim 0.8$ and $\mu \gtrsim 1 \tev$, giving largest scattering cross-sections around $\sigma_{\rm SI}  \lesssim 6 \times 10^{-45} \text{cm}^2$. However, as discussed in Section 6.3, the constraints cannot all be satisfied simultaneously, so in practice only a cross section of about $10^{-45} \cm^2$ could be achieved.

\end{document}